# MICROPATTERN GASEOUS DETECTORS .

TOM FRANCKE
*XCounter AB, Danderyd,  18233, Sweden*

VLADIMIR PESKOV
*Particle Physics Group, Royal Institute of Technology,
Stockholm, 10691, Sweden*

1.  **Introduction**

   In the last century several major inventions in the field of gaseous detectors were made. Let us just mention the Geiger counter [1], parallel plate detectors [2] and multiwire proportional chambers (MWPC) [3]. The main feature of all of these detectors is that they exploit gas multiplication or Townsend avalanches. One of the most successful developments was the MWPC which combined the gas multiplication feature with a position resolution (typically 0.1x2 mm). The inventor of this detector -G. Charpak -was awarded the Nobel Price in 1992. Parallel to these main developments there have always been many small- scale efforts in the development of high granularity gaseous detectors or, to be more precise, gaseous detectors with small distances between the anode and the cathode electrodes. For example, very small single wire detectors were developed for medical applications [4], an  array of wire detectors was used for cosmic applications [5], small gap multiwire detectors were also developed for plasma diagnostics (see for example [6,7]. Such detectors may potentially offer high 2D position resolutions. However, the manufacturing techniques of such detectors were very difficult and they did not receive great attention.
A real breakthrough in this direction was made by A. Oed who suggested the use of microelectronic techniques for the manufacturing of gaseous detectors which makes the manufacturing a lot easier [8]. This triggered a chain of other inventions: microgap [9], MICROMEGAS [10] and GEM [11] to name a few (see below for more details) . Now all of these new detectors are called - micropattern gaseous detectors. As one can see from the following chapters, it is quite a wide class of detectors: from strips, to dots, to hole -type structures. It will thus be useful to introduce a definition for micropattern detectors: they are high granularity gaseous detectors with small (below 1 mm) distances between the anode and the cathode electrodes.

The main advantages of the micropattern detectors are that: new (microelectronics) technology was applied for their manufacturing and they have high "granularity", thus offering potentials for very high position resolutions. Moreover, due to this feature (small distance between the electrodes) they may have high time resolutions and good counting rate capabilities.

The aim of this report is to review the main achievements in this field and identify any possible future progress whilst describing new and current applications.

## 2. Main Directions in the Design of Micropattern Gaseous Detectors

More than 20 various designs of micropattern detectors are known. Most of them are already described in several review papers [12,13] so to avoid repetition we will just mention the main designs and then focus our attention on the description of the main tendencies in the developments.

### *2.1 Microstrip (Microwire) –Type of Gaseous Detectors*

A classical example of a microstrip detector is one invented by A. Oed [8]. It is an alternative cathode and anode strips structure (typical pitch of 200- 400 µm) deposit by a lithographic method on a dielectric supporting structure (see Fig. 1). The small thickness of the anode strips (7-20 µm) ensures the formation of a high electric field in its vicinity. Primary electrons created by external radiation in the volume between the drift electrode and the anode-cathode plane (typically the thickness of this gap is L= 3-30 mm) drift toward the anode strips and trigger Townsend avalanches. Typical gains which could be achieved were around $10^4$. There are several variations of this main design however, for example the microgap gas counters [9] and the WELL detector [14].

In most cases there are 2D multiplication structures but recently 3D "strip" or "microwire" structures were developed as well [15].

### *2.2 Microdot (Micropin) –Type Detectors*

A basic microdot detector is a periodic structure of coaxial cathode and anode rings deposited by lithographic technology on a dielectric substrate-see Fig. 2 [16, 17].

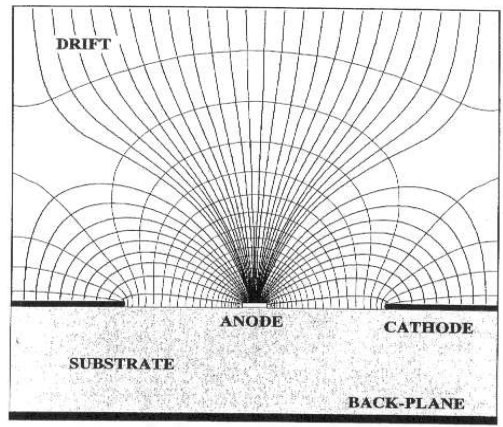

Fig. 1. A schematic drawing of a microstrip gas counter [18].

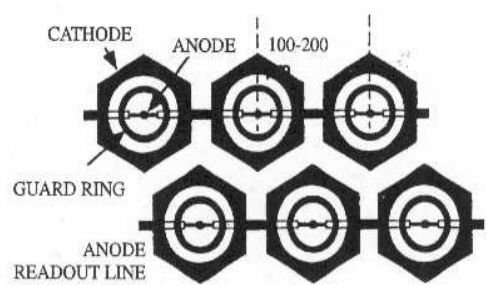

Fig. 2. Schematics of a microdot chamber [18].

Typical diameters of the cathode and the anode rings are 200 and 20 $\mu$m respectively. As in the previous detector, the absorption of the radiation occurs in the gas volume between the drifting electrode and the anode-cathode electrode's plane (L=3-30 mm). Due to the small diameter of the anode dots the electric field lines are focused in their vicinity. Primary electrons created in the drift volume by the external radiation drift toward the anode dots and initiate Townsend avalanches. Typical gains of microdot detectors are $10^4$, however in some optimised gas mixtures gains of up to $10^5$ were possible to achieve.

There are several variations to this basic design. Recently, 3D versions of microdot detectors were developed by several groups [19,20]. This detector, which in fact is resembling an array of micro single-wire counters (see [6]), had invertors expectations that very high gains, typical for single wire detectors, would be achieved [21], however, with the present designs gains of only ~$10^4$ were possible to reach.

### *2.3 Hole –Type Detectors*

Typically, the hole –type structure is a metallized from both sides dielectric sheet (typically 0.05-2 mm thick) with holes (0.1-2 mm in diameter). If a high voltage is applied to the metallic electrodes then each individual hole works as an electrostatic lance focusing field lines. This allows a high electric field inside the holes to be formed. Primary electrons created by external radiation in the drift region (the space between the drift electrode and the hole-type multiplication structure) drift to the holes and trigger Townsend avalanches in that area.

Most likely the first authors who demonstrated that some gain is possible to achieve inside the glass capillaries (with inner diameters of 0.1-5 mm) were A. Del Guerra et al [22]. However, this idea recently gained a new momentum after the suggestions of several authors in using micro-holes [23, 24] and "trench" – type structures [25]. The most popular one today is a so called Gas Electron Multiplier (GEM), developed by F. Sauli [11]. It is a metallized (from both sides) kapton sheet (50 μm thick) with holes of 100 μm in diameter and a pitch of 140 μm-see Fig. 3. GEM has several important advantages over other hole – type detectors: kapton is low in mass and is flexible as a material and such a detector is easy to manufacture for a low price. Gains of $10^4$ are possible to achieve with this detector. However nowadays in most applications, stacks of GEMs are usually used (like in traditional multistep avalanche chambers [26]). Such a multistep detector contains several GEMs placed 0.5-3 mm above each other. Voltages over each GEM and between the GEMs are set in a such a way that part of the multiplied charge in the holes could be extracted from the holes and directed to another GEM; so that cascade multiplication occurs (see [27] for more details). This allows the overall maximum gain to be boosted up to $10^6$.

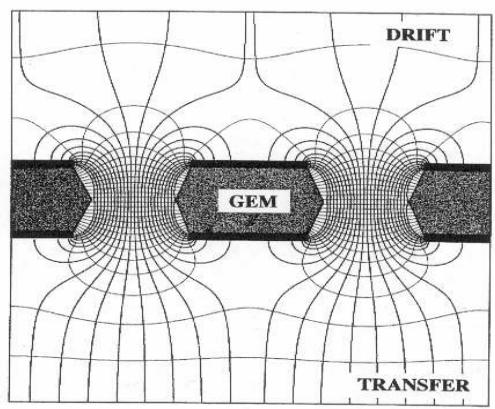

Fig. 3. A schematic drawing of the GEM. Field lines and equi-potentials are shown as well [18].

*2.4 Parallel-Plate Type Detectors*

A "classical" example of a parallel plate -type micropattern detectors is a so called Micro Mesh Gas Detector (MICROMEGAS) [10]. The main element in this detector is the micromesh (4-29 μm in wire's pitch) stretched 50- 100 μm above the readout plate (usually a G10 plate with metallic strips of a 300 μm pitch). A voltage of 400-700 V is applied between the mesh and the readout plate. The primary electrons created in the drift space (L=3-30mm) move toward the micromesh, drift through the mesh openings and then experience multiplication in the gap between the mesh and the readout plate. Typical gains are $\sim 10^5$ (at low counting rates - see paragraph 7). One can see that this detector is similar to parallel-mesh detectors widely used previously (see for example [28]). However, the revolutional step was that the gap between the mesh and the anode plate was reduced by almost two orders of magnitude.

This idea of a microgap parallel- plate detector triggered a chain of other inventions. One of them are the microgap Resistive Plate Chambers (RPCs), which immediately began to be used in practice. There are two main developments in this direction: a "timing RPC" [29] and a high position resolution RPC [30, 31]. "Timing" RPCs are parallel plate detectors with the metallic cathodes and anodes made of medium resistivity ($\rho \sim 10^9$-$10^{11}$ Ωcm) glass. The gap between the cathode and the anode is typically 100-400 μm [29]. The small gap allows one to achieve a very high time resolution, ~50 ps [29, 30]. The main applications today for this RPC are time of flight detectors for high-energy physic and PET [30-32].

High position resolution RPCs have a slightly different design: their cathodes are mode of low resitivity ($\rho \sim 10^4$-$10^8$ $\Omega$cm) materials (Si, GaAs) and the anodes –from medium resistivity glasses with metallic strips of 50 μm pitch-see Fig.4.

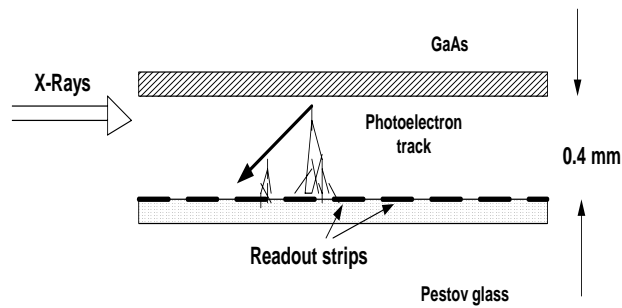

Fig. 4. A schematic drawing of a high position resolution RPC.

The gap between the anode and the cathode is about 100-400 μm. This allows one to achieve an excellent position resolution- better than 50 μm in digital mode [30]. It is remarkable that such RPCs can operate almost at the same gas gains and counting rates as metallic parallel plate avalanche chambers (PPACs) [30]. But in contrast to the metallic PPAC they are spark protected and thus very reliable in exploitation. This is why these detectors almost immediately after their development began to be used for medical imaging applications [33].

Recently, a microgap detector design combining both ideas (MICROMEGAS and microgap RPCs) has been reported [34]. It has a mesh cathode and a medium resistivity anode placed 100 μm below the mesh. The main advantages of this detector are that it is spark- protected and has a traditional drift volume (actually this design is similar to one the described in [35], but with a much smaller gap between the cathode´s mesh and the anode plate).

## 3. Signal Readout Techniques

There are two main techniques for signal readout from micropattern detectors: the use of the induced charge from the metallic readout strips or pads is one technique, and the use of the light emission produced by Townsend avalanches is another. In the case of the microstrip detectors (see Fig. 1), the charge - sensitive amplifiers are usually directly connected to the anode or the cathodes strips. In addition, induced signals from the backside of the dielectric supporting plane are used quite often. The readout strips on the back plane are usually oriented perpendicularly to the anode and the cathode strips. This allows one to obtain 2D images of the detected events.

In the case of the hole- type detector a readout plate is usually placed ~1mm below the detector. To obtain 2D images either a system of strips isolated from each other and oriented perpendicular to each other or pads are used [12,36]. Recently, very promising results were obtained with a so-called "active pixel" readout – an amorphous silicon thin-film transistors array [37].

In the case of the parallel-plate micropattern detectors, the anode plate serves as a readout plate simultaniously. The anode plate may have metallic readout strips both in the inner and outer surfaces.

The optical readout of gaseous detectors with a TV tube or a CCD camera has been used for a long time now [38, 39]. However, the application of it to the micropattern detectors gave a new momentum to this technique, allowing one to obtain impressive images of various objects or particle's tracks [40, 41, 42].

## 4. Efforts in the Optimization of the Micropattern Detector's Design

As one can see from a short review presented above, most micropattern gaseous detectors have maximum achievable gains of $A_{max}=10^4$-$10^5$. Note that this is 10-100 times less than is possible to reach with usual gaseous detectors, wire or parallel-plate type. However, even these relatively moderate gains were achieved after careful studies by many authors on the detector's design optimisation [43-46]. Thus one can consider today gains of $10^4$-$10^5$ as the maximum that can be achieved for micropattern gaseous detectors. The discussion for the reasons of these limits is given in paragraph 7.

## 5. Main Tendencies in the Developments Today

One can identify two very natural directions in the development of micropattern gaseous detectors today: 1) inventions of new designs and technology for their manufacturing, 2) improvement of their reliability in the existing devices.

Main tendencies in the development of new designs are: attempts to reach the highest possible granularity (or the smallest possible distance between the anode and cathode structures) [9,23, 47], attempts to restrict the released energy in case of occasional discharges by using for example, resistive materials for electrodes [33] and developing 3D multiplication structures (see for example [15,20]).

The main tendencies in improving the existing devices are: an attempt to increase the maxim achievable gain and attempts to make their operation more reliable by avoiding discharges or making them less harmful [33,49].

The limit in maximum achievable gain mentioned in the previous section creates serious problems in real applications. Indeed, gains of $10^4$-$10^5$ are the maximum that can be achieved. In the presence of heavily ionized particles or at high enough counting rate the maximum achievable gain further drops. Thus to guard the detector from possible destructive sparks, one has to operate at gains of only ~$10^3$.

There are two ways to overcome this limit: either operating at gains close to the $A_{max}$, but insuring at the same time that discharges, if they appear, are harmless or by using micropattern detectors in combination with some preamplification structure.

The first approach is more appropriate for parallel plate type micropattern detectors because they have the highest value of $A_{max}$ compared to other micropattern detectors [49,50,43]. To protect this type of detector against destructive sparks either resistors are used connected to each individual readout strip (as was in the case with MICROMEGAS [49]), or their electrodes are made of resistive materials (as was in the case with the microgap RPCs [30]). This allows the operation at gains close to $10^5$ to be at low (<10 Hz/mm$^2$) counting rates. As was mentioned before, at higher counting rates or in the presence of heavily ionized particles, the maximum achievable gain drops (see [50] for more details).

The second approach is now mostly used for all other types of micropattern detectors, for example in microstrip or hole -type. As a preamplification structure either a parallel mesh detector [52], GEM [53], or a capillary plate [54] can be used. The exact type of preamplifiaction structure is dictated by practical requirements only. From the point of view of physics they are all equal. The main idea in the use of the preamplification structure is to reduce the gain in each multiplication stage. The reason why this allows one to reach overall high gains will be discussed in paragraph 7.

## 6. How Far Can We Go?

As was mentioned above, one of the tendencies in the development of micropattern gaseous detectors is the attempt to achieve the highest possible granularity (or the smallest possible distance Δ between the anode's and cathode's electrodes). "Micro"-micropattern detectors with $\Delta \sim 1.5$ μm were already tested, some were successful [23, 48] - Fig. 5,6.

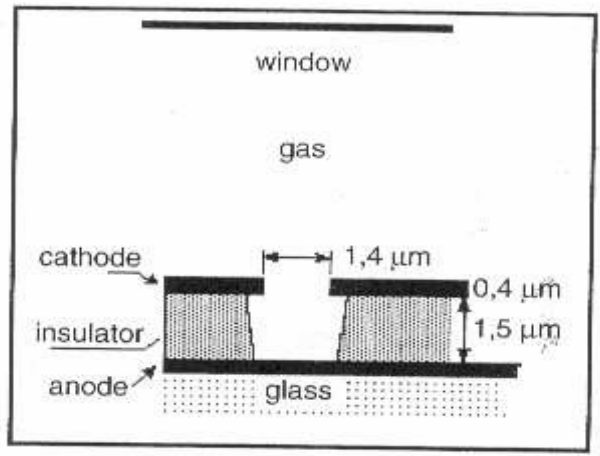

Fig. 5. Electrode structure based on a flat screen display [23].

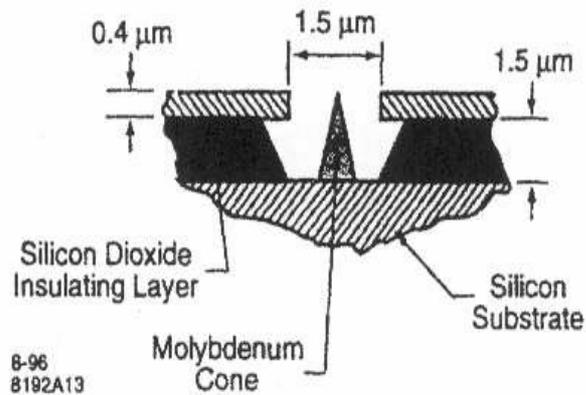

Fig. 6. The concept of the "Spind "ctathode detector [48].

One can ask the natural question: what sets the limit in these developments? Can we further decrease $\Delta$ and use for example, nanostructures?

Obviously for the moment there is no limit on the manufacturing technology of such structures. The limit is actually set by the working media- by gas in the given case. Experiments show that the maximum achievable gain for "micro"-micropattern detectors was small, ~30. The other concept associated with this problem is that any breakdown may easily destroy this fragile electrode structure. Only one discharge could be fatal; so the reliability of such detectors could be questionable.

Finally, can one really benefit in practice from very high granularity? Note that in many cases the range of delta electrons or photoelectrons is much larger than a few μm [55], so do we need such a high granularity or segmentation?

In the next chapter we will discuss what factors limit the maximum achievable gain and the rate characteristics of micropattern detectors, as well as what size the gap between the electrodes should be in order to be sufficient in practice.

**7. Gain Limit**
*7.1 Low Rate*

There are at least two main phenomenas contributing to the gain limit ($A_{max}$ at which breakdowns appears) of micropattern gaseous detectors at low counting rates:
1. Streamers in gas volume and
2. Streamers across the dielectric surfaces.

Historically, systematic studies of streamers in gas were done by Raether in large gap (>3 mm) parallel plate avalanche chambers. He experimentally established that in this detector's geometry streamers might appear when the total charge in the avalanche exceeds some critical value [2]:

$A_{max}n_0 \geq Q_{max} \sim 10^8$ electrons  (1),

where $n_0$ is the number of primary electrons created by the radiation in the gas. Note that the maximum achievable gain $A_{max}$ is inversely proportional to $n_0$. Thus for 6 keV X-rays ($n_0 \sim 200$ electrons) the breakdown will appear at gains of $10^6$, whereas for alpha particles ($n_0 \sim 10^5$) the maximum achievable gain will be $A_{max} \sim 10^3$.

The value of $Q_{max}$ at $\sim 10^8$ electrons is often called the "Raether" limit. The physic behind this is that at $A_{max}n_0 \sim Q_{max}$ the space charge in the avalanche becomes sufficient to disturb the external electric filed. As a result, photoelectrons created by avalanches in the surrounding gas volume begin to drift toward the positive ions remaining from the initial avalanche and finally forming a thin plasma filament, called a streamer –see Fig. 7 and [2, 56].

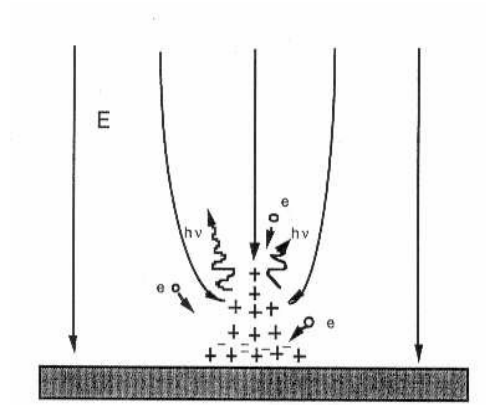

Fig. 7. Schematic drawing of streamer developments [56].

Note that the streamer causes breakdown when its head reaches the cathode. In the case of an almost uniform external electric field all streamers reach the cathode. However, if the field strength drops quickly with the distance from the anode's electrode, the streamer's propagation may stop in the gas volume without reaching the cathode [56]. These "self-quenched" streamers do not cause any harmful breakdowns.

What is described here is correct for usual gaseous detectors (parallel plate- type or wire- type detectors with thick anode wires). It was recently found [50,56] that in most micropattern gaseous detectors breakdowns also appear at some critical charge in avalanche :

$A_{max}n_0 \geq q_{max}$   (2),

where $q_{max}$ is some critical value ($q_{max} < Q_{max}$) which depends on the micropattern detector's geometry and the $n_0$ (see [56]). For example, in the case of parallel plate geometries it linearly increases with the thickness of the gap d:

$q_{max} \sim kd$   (3),

where k is a coefficient.

Thus at small gaps the breakdown will appear at a smaller total charge. Since this type of breakdown is associated with the space charge effect, it depends also on electron's density $n_e$ in the cloud of primary electrons $n_e \sim V_{no}/n_0$. The volume of the cloud $V_{no}$ in turn depends on the density of the gas and also on the diffusion process. For example, after multiplication in the GEM holes, the charge cloud expands and this explains why the pre-amplification structures allow one to reach overall high gains [52].

Note that there could also be other phenomena restricting the $A_{max}$ - electron jets being emitted from dielectric insertions on the cathodes of micropattern detectors [50,51,58]. Such insertion could be, for example: residues due to mechanical or chemical treatments, dust particles, dirt ect. Some vapors and gases can also form thin absorbed layers (actually liquid layers) on the cathode's surface which then play a role of the insertions. Such "insertions" could accumulate some positive surface charges due to ions from previous avalanches. This surface charge may create extremely high electric fields inside thin dielectric films and cause so called "explosive" field emission- jets of electrons sporadically distributed in time (see [50, 51, 58] for more details). The number of electrons in each emitted jet could stochastically vary between a few to up to $10^5$. These electrons in turn trigger Townsend avalanches in the micropattern detector. If at some moment the number of primary electrons in the jet satisfies the condition (2), then the streamer could be formed and a breakdown will appear.

The other important phenomena in the operation of micropattern gaseous detectors are surface streamers occurring across the dielectric supporting structures between the anode and the cathode's electrodes [56]. The formation of the surface streamers is not directly connected to the value of $q_{max}$. They could develop when avalanches along the surface begin to propagate and the electric field due to the avalanche's space charge and it's image in the dielectric reaches some critical value. In this case, photoelectrons from the surface or the surrounding gas begin to move towards the initial surface- attached avalanche and forming gliding discharges. Surface streamers may prevent one to reach the maximum possible gains determined by the condition (2). It could be a serious problem at low distances between the electrodes.

The works on the optimization of micropattern detector's designs mentioned in paragraph 4 were actually attempts to avoid conditions for the formation of streamers in the gas and across the dielectric surfaces. As was described above, the streamer in the gas could be suppressed if the external electric field drops sharply with the distance from the anode [56]. This is why micropattern detectors with thin anode strips (or with dot-type anodes) offered the highest gains [43].To restrict streamer propagation along the surfaces, specially shaped dielectric surfaces could be useful (for example -surfaces with grooves) [14,43].

*7.2 High Counting Rates*

It is a well established fact now that for all micropattern detectors the maximum achievable gain drops with the rate [50,51,58]-see Fig 8. It will be useful at this point to clarify a typical confusion. It has been known for a long time that for usual wire –type detectors the actual gain A drops with the rate H (due to the space charge effect) and this actually prevents sparking at high rates. In contrast, in the case of most micropattern detectors, the actual gain remains unchanged with the rate: for each chosen value of A the function A (H)=const. (see [58] for more details). However, the maximum achievable gain at which sparking appears $A_{max}$ drops with the rate-see Fig. 8. The confusion mentioned above come from the fact that function $A_{max}$ vs. rate for micropattern detectors looks very similar to the function A vs. rate for wire chambers. This formal similarity cause very common mistake in interpretation that in both cases this is the same effect: gain reduction due to the space charge created by avalanches. However it is not true (or not completely true). One can see from Fig. 8 that the $A_{max}$ vs. the rate curve has different slopes which may reflect different physical mechanisms responsible for breakdowns. Indeed, even at the rate range of $10^2$-$10^3$ Hz/mm$^2$, the maximum achievable gain already begins to drop. Let us consider for simplicity an example with MICROMEGAS (see [59]). It is known that the ion- removal time from the MICROMEGAS's gap is ~100 ns, so at these rates the positive ions for each particular avalanche are completely removed before the next avalanche begins to develop. Each avalanche therefore, develops completely independently from the previous one, so certainly there is some "memory" effect: the detector "remembers" for quite a long time the previous avalanches and this somehow affects the maximum achievable gain. As was shown in several studies, this "memory" effect could be the charging up of the dielectric layers (including absorbed liquid layers [51]) and insertions on the cathode's surfaces and associated with that jets of electrons.

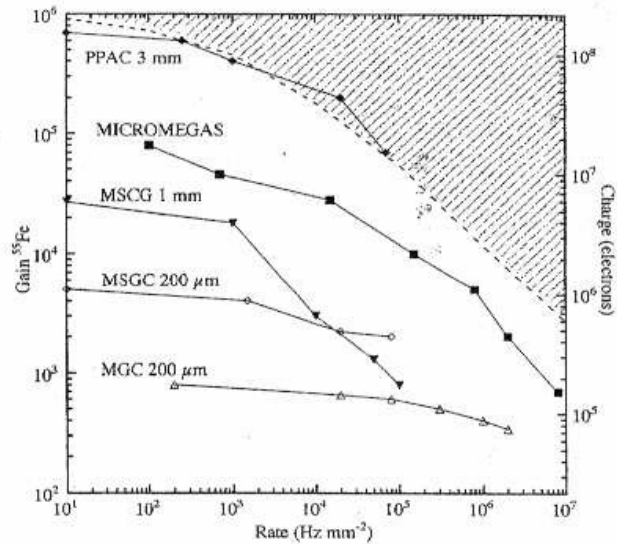

Fig. 8. Maximum gain vs. rate for several micropattern detectors. A dash line delimits the forbidden region (where discharges appear) [12, 50,51].

At higher rates other effects may contribute as well: statistical overlaps of neighbouring avalanches in time and space (so the condition (2) could be satisfied) [60], modification of the electric field in the cathode –anode's gap due to the steady space charge, multistep ionization [61,62], gas heating and accumulation of the exited atoms and molecules (which may also lead to a sudden current growth and breakdown (see [62] for more details).

*7.3 The Position Resolution*

In the case of the detection of charged particles a very high position resolution could be achieved with micropattern detectors. For example, in tracking measurements $\sigma \sim 12$ μm was achieved with MICROMEGAS [63] and ~40 μm with the GEM [64]. Thus the high granularity of micropattern detectors plays an important role in this application.

However, in the case of the detection of X-rays it is not a straightforward task to exploit the high granularity of micropattern detectors. Indeed, the range of photoelectrons even in the heavy gases could reach a few mm and the fluorescent photons can propagate and cause other ionization events in as far as 100-300 mm from the first absorption event [65]. Thus high granularity becomes quite useless. The standard approach in reaching a reasonable position resolution is by operating the detector at an elevated pressure. Recently however, an extremely high position resolution (better than 50 μm) was achieved with a microgap RPC operating at 1 atm [30]. The reason is clear by looking at Fig 4. If an X–ray radiation enters the detector close to its cathode and parallel to it, the photoelectron tracks will originate from this area. The unique feature of the parallel-plate's geometry is that the gas multiplication factor depends exponentially on the distance of the primary electrons from the cathode. As a result, the main contribution to the signal on the readout strips gives the primary electrons created near the cathode - a region where the collimated X-ray beam is introduced. The other part of the photoelectron track, even if it is very long but inclined (and most of the tracks are inclined), contributes very little to the signal amplitude. Thus in this particular geometry one can detect mostly the vertex of the photoelectron track which ensures extremely high position resolutions. In this particular method the high segmentation of the micropattern detector plays a crucial role.

We can now try to answer the question which we posed at the beginning of section 6: is there a need to develop "micro"-micropattern detectors with $\Delta$ of a few μm? For most applications today a position resolution of 12-40 μm is sufficient. Of course this fact will not stop further developments and certainly more and more new designs of various "micro"-micropattern detectors with position resolutions ~1 μm will appear in the nearest future. These designs may open new avenues in applications. One of these will be described in section 8.4.

## 8. Applications

Nowadays the main applications for the micropattern are the same as for usual gaseous detectors:
high energy physics,
astrophysics,
plasma diagnostics,
medicine,
biology and
industry.
These traditional applications of micropattern detectors have already been described in several review papers [12, 13, 29, 66]. For this reason, to avoid any repetition we will here focus only on very recent and very "exotic" applications:
the detection of visible photons,
operation at extremely high counting rates (up to $10^{10}$ Hz/mm$^2$),
UV and X-ray imaging with position resolution of 30-50 μm at counting rates of $10^5$ Hz/mm$^2$ and operation inside LAr/Xe.

### *8.1 The Detection of Visible Photons*

During the last decade or so there have been a lot of efforts in the development of gaseous detectors sensitive to visible light [67, 68]. The potential advantage of such detectors, compared to traditional vacuum ones, is their insensitivity to magnetic fields and their possibility of using large-area photocathodes at low costs. Unfortunately, traditional gaseous detectors combined with photocathodes sensitive to visible light suffer from feedback even at gains of 50 –100. This gain is too low to detect single photoelectrons. These developments gained a new momentum with hole- type micropattern detectors (capillaries and GEM). As was mentioned in paragraph 2.3 and in [67, 68], hole-type multipliers have two important advantages over the traditional avalanche detectors:
1. efficient reduction or suppression of ion and photon feedbacks; and

2. a possibility of charge extraction: primary electrons or avalanche–induced secondary electrons can be extracted from the holes and directed to a successive multiplication element.

In recent woks with multistep hole-type detectors combined with photcathodes, gains exceeding $10^3$ were successfully achieved [68]. The first attempts on manufacturing sealed prototypes were also made [69-71]. Nowadays, Hamamatsu is evaluating these types of photomultipliers as possible commercial products [72].

*8.2 Portal Imaging*

Radiation therapy to day is applied to approximately 50% of cancer patients. During treatment it is extremely important to monitor precisely the absolute intensity of the beam and its position with the respect to the tumor and the organs at risk. This could be done with a so called portal imaging device. It is designated for the monitoring and precise alignment of the pulsed cancer treatment gamma beam with respect to the patient's tumor position. The latest will be determined from an X-ray image of the patient obtained in the time intervals between the gamma pulses. During treatment, the image of the gamma beam profile can be compared to the X-ray image. The data could then be fed back to the treatment machine, making fast online corrections in the gamma beam position. Recently, a prototype of a simple and cheap electronic portal imaging device based on hole-type detectors, GEMs and capillaries, was developed and tested (see [73] for more details). It was demonstrated that the GEM and the capillaries could operate stably at extremely high counting rates: $10^7$-$10^{10}$ Hz/mm$^2$. The first images obtained with this prototype can be found in [74].

*8.3 High Counting Rate X-ray Imaging*

In the last decade enormous efforts by various research groups and companies have been made to develop digital radiographic devices. The most attractive among them are so-called "photon-counting devices", which allow one to reduce the dose during the image taking. The high–position resolution RPC described in paragraph 2.4 is one of the most promising candidates for this application [30]. As an example, Fig. 9 shows the image of a fish obtained with a high position resolution RPC and for comparison, the same image obtained by standard film techniques. It is obvious that the quality of the digital image obtained with the RPC is much higher. As was described earlier (see chapter 2.4) this device can operate at rates as high as the parallel-plate detectors with metallic electrodes. However, in contrast to the metallic PPAC it is spark-protected and has a position resolution better than 50 μm in digital mode at a counting rate of $10^5 Hz/mm^2$ [30].

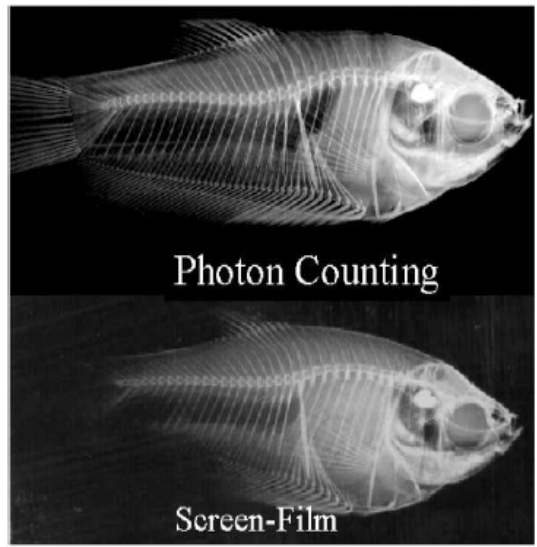

Fig. 9. X-ray images of the fish obtained with the high position resolution RPC and with a standard film [75].

## 8. 4 The Operation of Micropattern and "Micro"--micropattern detectors inside LAr/Xe

As was mentioned above, limits in gains and position resolutions of micropattern detectors are actually imposed by the working media –gas (see chapter 6). From this point of view it will be interesting to investigate their operation in a more dense media. Several experiments were made by various groups with micropattern detectors operating at high pressure gases and inside noble liquids. It has been discovered, for example, that micropattern detectors can operate inside noble liquids at gains of $>10^3$ [76,77]. The first observations of avalanche multiplication inside the noble liquids were done with "micro"-micropattern detector [76] as well as in the so-called "Spindt cathode" shown in Fig. 6 [76].

The potential advantage of liquids over gases for the detection of particles and X-rays is that the intrinsic position resolution in liquids is much better because the density of the ions in the tracks is much greater and because the diffusion of the drifting electrons is less. Successful operations of micropattern detectors inside noble liquids may open new avenues in applications, for example in noble liquid PETs [78] or WIPM detectors [79].

## 9. Conclusions

A revolution is currently taking place in the development of gaseous detectors of photons and particles. Parallel plate-type and wire-type detectors, which dominated for years in high energy and space flight experiments, are now being replaced by recently invented micropattern gaseous detectors. Since these detectors are cheap, can operate at relatively high gains and have very good position resolutions, they may compete with other types of detectors, for example with solid state detectors; especially in those fields of application where very low deposit energies are necessary to detect (tracking, X-rays, UV and visible photons) or where large sensitive areas are needed.